\documentclass[twocolumn,tighten]{aastex62}

\usepackage{xcolor}
\usepackage{hyperref}
\usepackage{afterpage}
\usepackage{xspace}
\usepackage{upgreek}
\usepackage{multirow}
\usepackage{xspace}

\usepackage{amsmath}

\usepackage{hyperref}
\usepackage[frozencache,cachedir=.]{minted}

\newcommand{\setigen}{\texttt{setigen}\xspace}
\newcommand{\Var}{\mathrm{Var}}

\newcommand{\subplotref}[2]{\hyperref[#1]{\ref{#1}#2}}

\graphicspath{{./}{figures/}}

\shorttitle{Setigen}
\shortauthors{Brzycki et al.}

\begin{document}

\title{\setigen: Simulating Radio Technosignatures for SETI}


\newcommand{\UCB}{Department of Astronomy, University of California Berkeley, Berkeley CA 94720}
\newcommand{\BL}{Breakthrough Listen, University of California Berkeley, Berkeley CA 94720}
\newcommand{\SSL}{Space Sciences Laboratory, University of California, Berkeley, Berkeley CA 94720}
\newcommand{\SWIN}{Centre for Astrophysics \& Supercomputing, Swinburne University of Technology, Hawthorn, VIC 3122, Australia}
\newcommand{\GBT}{Green Bank Observatory, West Virginia, 24944, USA}
\newcommand{\NIJ}{Department of Astrophysics/IMAPP,Radboud University, Nijmegen, Netherlands}
\newcommand{\OXF}{Astronomy Department, University of Oxford, Keble Rd, Oxford, OX13RH, United Kingdom}
\newcommand{\UMAN}{Department of Physics and Astronomy, University of Manchester, UK}
\newcommand{\ATNF}{Australia Telescope National Facility, CSIRO, PO Box 76, Epping, NSW 1710, Australia}
\newcommand{\HOU}{Hellenic Open University, School of Science \& Technology, Parodos Aristotelous, Perivola Patron, Greece}

 \newcommand{\USQ}{University of Southern Queensland, Toowoomba, QLD 4350, Australia}

\newcommand{\SETI}{SETI Institute, Mountain View, CA 94043}
\newcommand{\KZA}{University of Malta, Institute of Space Sciences and Astronomy, Malta}
\newcommand{\PWJD}{The Breakthrough Initiatives, NASA Research Park, Bld. 18, Moffett Field, CA, 94035, USA}

\newcommand{\UT}{Dunlap Institute for Astronomy \& Astrophysics, University of Toronto, 50 St.~George Street, Toronto, ON M5S 3H4, Canada}

\newcommand{\CU}{International Centre for Radio Astronomy Research, Curtin University, Bentley WA 6102, Australia}

\newcommand{\UR}{Goergen Institute for Data Science, University of Rochester, Rochester, NY 14627}

\newcommand{\newtext}[1]{\color{black} #1} 


\correspondingauthor{Bryan Brzycki}
\email{bbrzycki@berkeley.edu}

\author[0000-0002-7461-107X]{Bryan Brzycki}
\affiliation{\UCB}

\author[0000-0003-2828-7720]{Andrew P.\ V.\ Siemion}
\affiliation{\BL}
\affiliation{\SETI}
\affiliation{\UMAN}
\affiliation{\KZA}

\author[0000-0002-4278-3168]{Imke de Pater}
\affiliation{\UCB}

\author[0000-0003-4823-129X]{Steve Croft}
\affiliation{\UCB}
\affiliation{\SETI}

\author[0000-0001-5591-5927]{John Hoang}
\affiliation{\UCB}

\author[0000-0002-3616-5160]{Cherry Ng}
\affiliation{\UT}
\affiliation{\UCB}
\affiliation{\SETI}

\author[0000-0003-2783-1608]{Danny C. Price}
\affiliation{\CU}
\affiliation{\UCB}

\author[0000-0001-7057-4999]{Sofia Sheikh}
\affiliation{\UCB}

\author[0000-0003-3248-2174]{Zihe Zheng}
\affiliation{\UR}

\begin{abstract}

The goal of the search for extraterrestrial intelligence (SETI) is the detection of non-human technosignatures, such as technology-produced emission in radio observations. While many have speculated about the character of such technosignatures, radio SETI fundamentally involves searching for signals that not only have never been detected, but also have a vast range of potential morphologies. Given that we have not yet detected a radio SETI signal, we must make assumptions about their form to develop search algorithms. The lack of positive detections also makes it difficult to test these algorithms' inherent efficacy. To address these challenges, we present \setigen, a Python-based, open-source library for heuristic-based signal synthesis and injection for both spectrograms (dynamic spectra) and raw voltage data. \setigen facilitates the production of synthetic radio observations, interfaces with standard data products used extensively by the Breakthrough Listen project (BL), and focuses on providing a physically-motivated synthesis framework compatible with real observational data and associated search methods. We discuss the core routines of \setigen and present existing and future use cases in the development and evaluation of SETI search algorithms. 

\end{abstract}


\section{Introduction}
\label{sec:intro}

Since the inception of radio SETI in the 1960s, technosignature searches have greatly expanded to cover more sky area, wider frequency ranges, and a larger variety of signal morphologies \citep{Drake:1961, werthimer1985serendip, tarter2001search, Siemion:2013, wright2014near, MacMahon:2018, Price:2018, gajjar2021breakthrough}. Arguably the most developed branch of radio SETI is the search for narrow-band technosignatures, with signal bandwidths under 1\,kHz, for which search algorithms are constantly being produced and improved \citep{Siemion:2013, Enriquez:2017, pinchuk2019search, margot2021search}. These algorithms operate on either voltage time series data or time-frequency spectrogram data (i.e., dynamic spectra, waterfall plots).

The incoherent tree deDoppler method is the primary search strategy for Doppler-accelerated narrow-band signals in radio spectrograms \citep{Taylor:1974, Siemion:2013, Enriquez:2017, margot2021search}. An ideal sinusoidal emitter will appear to exhibit a frequency drift over time due to relative acceleration between the emitter and receiving telescope \citep{sheikh2019choosing}. Under a constant relative acceleration, such a signal will have a linear drift or slope in a spectrogram of Stokes I intensities. The tree deDoppler algorithm efficiently integrates spectra over potential drift rates and identifies signals above a threshold signal-to-noise ratio (SNR). Breakthrough Listen, the most comprehensive SETI search program to date \citep{Worden:2017}, developed \texttt{turboSETI}\footnote{\url{https://github.com/UCBerkeleySETI/turbo_seti}}, an open-source implementation of the deDoppler algorithm that serves as the backbone of many technosignature searches \citep{Enriquez:2017, enriquez2019turboseti, Price:2020, sheikh2020breakthrough, gajjar2021breakthrough}. 

This method works well for signals with high duty cycles and linear drift rates, but it can struggle to properly detect more complex signals \citep{pinchuk2019search}. This is particularly problematic given the increasingly complex radio frequency interference (RFI) environment within which these searches are conducted. Moreover, the lack of robust, labeled, narrow-band signal datasets can make it difficult to quantify a given implementation's detection accuracy, especially in light of RFI and variable bandpass responses.

For more complex signal morphologies, machine learning (ML) algorithms have been proposed that use computer vision techniques to classify image-like spectrograms. However, the same lack of labeled, narrow-band signal data makes creating supervised ML models difficult. \cite{Zhang:2019} used a self-supervised approach in which spectrogram data was divided in time into two halves, for which the ML task was to predict the second half given the first. For an ML-based direction-of-origin filter, \cite{pinchuk2021machine} used a separate non-ML method to detect signals and create an algorithmically-labeled spectrogram dataset. In most cases, however, supervised approaches have relied on generating synthetic signals of various classes in order to guarantee correct labels \citep{harp2019machine, brzycki2020narrow, margot2021search}. 

To address these issues, we present \setigen, an open-source Python library that facilitates the creation of synthetic narrow-band signals and supports injection into observational data. \setigen is meant to provide a general-use heuristic framework for creating mock radio SETI data. A primary design aspect is ensuring that the synthesis process is grounded as much as possible in physical quantities to better interface with real observations and search algorithms. \setigen makes heavy use of \texttt{NumPy}\footnote{\url{https://numpy.org/}} for efficient matrix operations \citep{oliphant2006guide, harris2020array} and \texttt{blimpy}\footnote{\url{https://github.com/UCBerkeleySETI/blimpy}} for interfacing with data products routinely used by BL \citep{2019JOSS....4.1554P}. 

There are two main modules in \setigen, ``spectrogram'' and ``voltage,'' dedicated to the most common data formats used in radio SETI. The spectrogram module works with Stokes I (intensity) data stored as time-frequency arrays and is designed to be flexible and heuristic-based. It can be used to generate many small snippets of data containing synthetic signals for quick algorithm test cases or for full labeled datasets. The voltage module creates synthetic antenna voltages, follows these voltages through a software-based signal processing chain that models a standard single dish signal pipeline, including quantization and a polyphase filterbank, and saves the final complex voltages. This requires a lot more computational power, so voltage \setigen routines can be optionally GPU-accelerated via \texttt{CuPy}\footnote{\url{https://cupy.dev/}} \citep{cupy_learningsys2017}. Since the voltage module models the signal processing chain, it can be used to produce more ``realistic'' signals, test complex voltage processing software, and evaluate how each signal processing element affects the final signal sensitivity. 

Radio SETI searches typically operate on data in spectrogram format, since it compresses data and enables visualization and analysis of broader signal morphology in time-frequency space \citep{Enriquez:2017, Margot:2018, Pinchuk:2019, Price:2020, sheikh2020breakthrough}. As such, \setigen was initially written to create large datasets of radio spectrograms for use in supervised ML search experiments. The library was later expanded to support synthesizing raw voltage-level data to complement existing use cases. 

\setigen has already been used in a variety of applications, such as the development and testing of search algorithms. It has been used to create synthetic datasets with position labels for ML localization tasks in single observations \citep{brzycki2020narrow}. \setigen has also been used to inject synthetic signals within ON-OFF cadences, each comprised of 6 consecutive observations and used as a direction-of-origin filter for SETI. Ma et al. (submitted) injected signals into ON-OFF cadences taken with the Robert C. Byrd Green Bank Telescope \citep[GBT;][]{MacMahon:2018} to train a sophisticated variational autoencoder model that can classify cadences as potential SETI candidates. Similarly, \setigen was used extensively to produce training and test data in BL's first Kaggle ML competition\footnote{\url{https://www.kaggle.com/c/seti-breakthrough-listen}}, in which contestants were tasked with classifying synthetic technosignature candidates in ON-OFF cadences. 

Outside of ML, synthetic \setigen data is used in injection-recovery testing for \texttt{turboSETI} as well as for a new search code, \texttt{hyperseti}\footnote{\url{https://github.com/UCBerkeleySETI/hyperseti}}. The voltage module has been used to test and upgrade parts of the Allen Telescope Array's \citep{welch2009allen} software signal processing pipeline. Furthermore, \setigen has been used to test RFI rejection and detection techniques for the Parkes Multibeam Galactic Plane Survey SETI search, helping to discriminate terrestrial signals from different regions in the sky as SETI surveys with multiple antennas or beams become more popular (Perez et al., in prep).

This paper is organized as follows. Section \ref{sec:signalchain} outlines the standard signal chain and processing pipeline used in single dish radio SETI observations to motivate details behind \setigen's synthesis methods. Section \ref{sec:methods} presents the code methodology: Section \ref{subsec:spectrogram} describes the spectrogram module for producing and working with synthetic Stokes I time-frequency data, while Section \ref{subsec:voltage} describes the voltage synthesis module in detail, connecting components of typical radio signal chains to software analogues used in \setigen. In Section \ref{sec:discussion}, we discuss current limitations of the library and future directions for signal synthesis for SETI.

\section{Overview of Single Dish Signal Chains}
\label{sec:signalchain}

To motivate the capabilities of \setigen, we first give a broad overview of the standard single dish data recording pipeline, as well as some details pertinent to the Breakthrough Listen digital recorder (BL DR) system at the GBT \citep[][]{MacMahon:2018}. 

In a single-dish radio telescope, incoming radiation is reflected off the dish surface toward a feed horn at the focus. The feed couples incident free-space electromagnetic radiation to voltages within the telescope's receiver system.

These voltages are passed to an analog down-conversion system containing a heterodyne mixer, which shifts the signal from the target RF range into an intermediate frequency (IF) range near baseband more suitable for receiver hardware. The resulting voltages are then digitized by analog-digital converters (ADC) to a specified number of bits $N_{\text{bits,d}}$ at a given sampling rate $f_s$. The BL DR system digitizes voltages to 8-bit at a sampling rate of $f_s = 3$ GHz for each linear polarization \citep{MacMahon:2018}. 

Radio telescope pipelines commonly use polyphase filterbanks \citep[PFB;][]{bellanger1976digital, harris2011mathematical, price2021spectrometers} to help partition the usable band and improve the spectral channel response of the system. For example, the BL DR system uses an 8-tap PFB to divide the 1.5\,GHz Nyquist range into $N_{\text{coarse}}=512$ ``coarse'' spectral channels, which in turn are divided among 8 compute nodes \citep{MacMahon:2018}. This procedure performs a Fast Fourier Transform (FFT) with a length of $P=2N_{\text{coarse}}=1024$. For receivers with wide bandwidths, such as C-band at 3.95--8.00\,GHz, multiple copies of these elements, starting from the analog mixer, are employed to cover the full band \citep{gbtpropguide}.

The digital processing components of the BL DR system are done on custom signal processing boards using field-programmable gate arrays (FPGAs), provided by the Collaboration for Astronomy Signal Processing and Electronics Research \citep[CASPER;][]{hickish2016decade}. These boards use fixed point arithmetic and increase numerical bit size when doing computations \citep{MacMahon:2018}. Accordingly, both real and imaginary components of the resulting complex voltages must be requantized (e.g. to $N_{\text{bits,r}}$) before they are written to disk. The BL DR system records these as 8-bit signed integers in GUPPI \citep[Green Bank Ultimate Pulsar Processing Instrument;][]{duplain2008launching} raw format, based on FITS \citep{pence2010definition} and stored as \texttt{.raw} files \citep{Lebofsky:2019}.

Since raw voltage data comes at the highest resolution possible given the ADC sampling rate, data volumes are large, especially during standard BL observing campaigns. Therefore, we finely channelize or ``reduce'' raw data into spectrograms (also known as dynamic spectra or ``waterfall plots''), 2D arrays of intensity (Stokes I) as a function of time and frequency \citep{Lebofsky:2019}. Multiple versions with different resolutions can be created from the same set of raw data by varying the FFT length $N_{\text{fine}}$ and integration factor $N_{\text{int}}$. 

During fine channelization, an FFT of length $N_{\text{fine}}$ is performed on complex raw voltages within individual coarse channels, resulting in $N_{\text{fine}}$ fine channels each. So, we can express the full Nyquist bandwidth as
\begin{equation}
    f_N = \frac{f_s}{2} = N_{\text{coarse}} N_{\text{fine}} \Delta f.
\end{equation}
This gives us an expression for the spectrogram's frequency resolution:
\begin{equation} \label{eq:df}
    \Delta f = \frac{f_s / 2}{N_{\text{coarse}} N_{\text{fine}}}.
\end{equation}

If the total observation length is $\tau$ and the number of time channels (pixels) in the final spectrogram is $N_t$, then
\begin{equation} \label{eq:nt}
    N_t = \frac{\tau}{\Delta t},
\end{equation}
assuming that $\tau$ is a multiple of the spectrogram's time resolution $\Delta t$. In practice, extraneous samples are truncated when necessary to satisfy this requirement.

The integration factor $N_{\text{int}}$ is the number of spectra integrated in the time direction. To get an expression for $\Delta t$, we can think in terms of the total number of samples collected (for a single linear polarization):
\begin{equation} \label{eq:ns1}
    N_s = \tau f_s.
\end{equation}
The pipeline takes in $N_s$ real samples in time and, via a $P$-point FFT, transforms the data into a complex 2D array in time-frequency space, with non-integrated dimensions $N_t N_{\text{int}} \times P N_{\text{fine}}$. 
\begin{equation} \label{eq:ns2}
    N_s = N_t N_{\text{int}} P N_{\text{fine}} = 2 N_t N_{\text{int}} N_{\text{coarse}} N_{\text{fine}}.
\end{equation}
Note that since the FFT is performed on real voltages, the unique frequency extent is ultimately halved per the Nyquist range.

Combining Eqs. \ref{eq:df}--\ref{eq:ns2}, we get
\begin{align}
    N_s = \tau f_s &= 2 N_t N_{\text{int}} N_{\text{coarse}} N_{\text{fine}} \\
    &= 2 \frac{\tau}{\Delta t} N_{\text{int}} N_{\text{coarse}} N_{\text{fine}} \\
    \Delta t &= \frac{2 N_{\text{int}} N_{\text{coarse}} N_{\text{fine}}}{f_s} = \frac{N_{\text{int}}}{\Delta f}. \label{eq:dt}
\end{align}
Although $N_{\text{fine}}$ and $N_{\text{int}}$ must both be integers, we otherwise have fine control over $\Delta f$ and $\Delta t$ through Eqs. \ref{eq:df} and \ref{eq:dt}.

\section{Code Methodology}
\label{sec:methods}
 
As object-oriented software, \setigen has a set of important classes and routines that are described below. For more technical details and examples of the API, see the full documentation\footnote{\url{https://setigen.readthedocs.io/}}.

\subsection{Spectrogram Module}
\label{subsec:spectrogram}

The spectrogram module provides an interface for synthesizing Stokes I (waterfall) data in a format common to radio SETI and is oriented around the \texttt{Frame} class. A \texttt{Frame} object contains a 2D data array of intensities as a function of time and frequency, as well as accompanying metadata, such as starting frequency and time-frequency resolutions. 

Data frames can be initialized from either saved observational data or frame parameters. Frames can extract Stokes I data and observational metadata from filterbank (\texttt{.fil}) or HDF5 files (\texttt{.h5}). The most important metadata for \setigen are the physical parameters of the underlying intensity data: resolutions and ranges in both time and frequency. Empty frames can therefore be created simply by specifying these parameters along with desired data array dimensions.

\subsubsection{Noise Synthesis}
\label{subsubsec:noise}

In most SETI applications, we search for statistically-significant signals embedded in noise. Since voltage noise in the absence of RFI approximately follows a zero-mean normal distribution \citep{ThompsonMoranSwenson}, the radiometer noise in spectrogram data follows a chi-squared distribution \citep{mcdonough1995detection, nita2007radio}. When the time and frequency resolutions are coarse enough, the spectrogram noise approaches a normal distribution by the central limit theorem.

Specifically, suppose we have a sequence of input voltages $\{x_n\}$ following a Gaussian distribution with zero mean. During the coarse channelization process, the polyphase filterbank applies, at its core, an FFT to bring the voltages into frequency space:
\begin{equation} 
    X_k = \sum_{n=0}^{N-1} {w_n x_n e^{-2\pi ikn/N}},\ k=0,\dots,N-1,
\end{equation}
where $N$ is the number of frequency bins and $\{w_n\}$ are coefficients of a windowing function applied to improve the spectral response \citep{price2021spectrometers}. 

More specifically, the filterbank sums over $M$ rows of $P$ samples before a $P$-point FFT, so that the response of the $r$th row of $P$ samples is:
\begin{equation} \label{eq:pfb}
    X_{k,r} = \sum_{p=0}^{P-1} \left[ \sum_{m=0}^{M-1} {w_{n'} x_{n''}} \right ] e^{-2\pi i k p / P},
\end{equation}
where $n'=mP+p$ and $n''=(r-M+m)P+p$ are indices of the windowing coefficients and voltages samples in terms of $m$ and $p$. Here, we assume that the $MP$ windowing coefficients are symmetric about the midpoint, so that $w_n = w_{MP-n-1}$.

Ignoring quantization for the moment, we store the complex components of the resulting FFT voltages, $\operatorname{Re}(X_k)$ and $\operatorname{Im}(X_k)$, as raw voltage data. Since these are linear combinations of independent zero-mean Gaussian variables (i.e. $x_n$), they both follow zero-mean Gaussian distributions. 

In the absence of a windowing function ($w_n=1$), for each channel besides the real-valued DC and Nyquist bins, the variances of the real and imaginary components are equal \citep[$\sigma^2$;][]{mcdonough1995detection}. When a windowing function is used, the underlying statistics can change such that the variances of the complex components differ as a function of spectral bin \citep{nita2007radio}. However, for commonly chosen symmetrical windows (e.g. Hamming), this effect is negligible in most spectral bins. 

For a single linear polarization, the power is given by
\begin{equation}
    I_{x,k} = |X_k|^2 = \operatorname{Re}(X_k)^2 + \operatorname{Im}(X_k)^2 
\end{equation}
Assuming both complex components have the same variance $\sigma^2$, the power follows a chi-squared distribution with two degrees of freedom:
\begin{equation}
    I_{x,k} \sim \sigma^2 \chi^2(2)
\end{equation}

During the fine channelization step, we integrate $N_{\text{int}}$ spectra in the time direction and combine power from $N_\text{pol}$ polarizations. Therefore, in the final Stokes I spectrogram, the total number of chi-squared degrees of freedom is given by:
\begin{align}
    \text{DOF} &= 2 N_{\text{pol}} N_{\text{int}} = 2 N_{\text{pol}} \Delta f \Delta t \\
    I_k &\sim \sigma^2 \chi^2(2 N_{\text{pol}} \Delta f \Delta t), \label{eq:ik}
\end{align}
using Eq. \ref{eq:dt}. For dual-polarization Stokes I data, $\text{DOF}=4 \Delta f \Delta t$. This allows us to generate synthetic chi-squared noise with the correct number of degrees of freedom just from frame resolutions, which are either directly specified or inferred from observations. Since non-calibrated intensity values are arbitrarily scaled, we can simply scale the magnitudes of synthetic chi-squared noise to match empirical observational noise distributions. 

The main function for noise synthesis across a frame is \texttt{add\_noise}, which adds random noise to every pixel in the data array. By default, it generates chi-squared noise with a user-specified mean intensity $\mu$. Since the mean of a chi-squared distribution equals the number of degrees of freedom, for dual-polarization data, we have
\begin{align}
    I_k &\sim \left( \frac{\mu}{4 \Delta f \Delta t} \right) \chi^2(4 \Delta f \Delta t) \label{eq:ik_mu} \\ 
    \langle I_k \rangle &= \left( \frac{\mu}{4 \Delta f \Delta t} \right) \cdot 4 \Delta f \Delta t = \mu \\
    \Var(I_k) &= \left( \frac{\mu}{4 \Delta f \Delta t} \right)^2 \cdot 2\cdot 4 \Delta f \Delta t = \frac{\mu^2}{2 \Delta f \Delta t}.
\end{align}

In addition to chi-squared noise, \texttt{add\_noise} can also generate Gaussian noise. By the central limit theorem, as the degrees of freedom increase, a chi-squared distribution approaches a normal distribution. For example, $N_{\text{int}}=51$ for BL's standard high spectral resolution data product, so $\text{DOF}=204$ and the resulting background noise is close to Gaussian. Directly synthesizing Gaussian-distributed noise can save normalization steps in data processing, but should be used carefully when comparing with real observational data.

A useful extension of the noise synthesis function is \texttt{add\_noise\_from\_obs}, which draws from archived observational statistics to set realistic intensity values. The observations were taken using the GBT at C-band and reduced to (1.4\,s, 1.4\,Hz) resolution. For example, for chi-squared noise, the function randomly selects an archived mean intensity, scales it to the appropriate frame resolution, and populates noise per Eq. \ref{eq:ik_mu}. An implementation detail of BL's fine channelization software, \texttt{rawspec}\footnote{\url{https://github.com/UCBerkeleySETI/rawspec}}, is that as part of the FFT, intensity values are scaled up by a factor of the FFT length $N_{\text{fine}}$. So, for observations going through the BL data pipeline (i.e. the same digitization and coarse channelization hardware):
\begin{align}
    \mu&\propto N_{\text{fine}} N_{\text{int}} \\
    &\propto N_{\text{fine}} \Delta f \Delta t \\
    &\propto \Delta t.
\end{align}
Alternatively, the function also accepts user-provided arrays of background noise intensity statistics from which to sample instead. This can be used for synthesizing data with intensity ranges from other telescopes (e.g. Parkes) or even GBT data at different frequency bands or sensitivities. 

After noise synthesis, the frame will update class attributes storing the estimated mean $\mu_b$ and standard deviation $\sigma_b$ of the background noise. For an empty frame, the first noise synthesis function will set these properties directly. For pre-loaded observational data and further noise injection, the frame estimates the background noise through iterative sigma clipping at the $3\sigma$ level to exclude outliers. For frames small enough that noise statistics do not change over the frequency bandwidth, this enables signal injection at desired SNR levels.

\subsubsection{Signal Synthesis}
\label{subsubsec:signal}

For narrow-band signal synthesis, the \texttt{add\_signal} function creates heuristic, user-defined signals in spectrogram data. Our convention is that the spectrogram data has time on the $y$-axis and frequency on the $x$-axis.

In spectrogram \setigen, narrow-band signals have a ``central" frequency at each timestep and a unique spectral profile centered at that frequency. As such, there are four main heuristic descriptors for a narrow-band signal in \setigen:
\begin{enumerate}
    \item \texttt{path} -- $I_p(t)$: Central signal frequencies as a function of time, e.g. linear (constant) drift rate, quadratic drift rate
    \item \texttt{t\_profile} -- $I_t(t)$: Signal intensity as a function of time, e.g. constant intensity, Gaussian pulses
    \item \texttt{f\_profile} -- $I_f(f, f_0)$: Spectral profile as a function of frequency (offset from central frequency), e.g. $\text{sinc}^2$ profile, Gaussian profile
    \item \texttt{bp\_profile} -- $I_{bp}(f)$: Bandpass profile as a function of absolute frequency
\end{enumerate}
These descriptors are parameters for \texttt{add\_signal} and are Python functions by type. A set of common functions are provided with \setigen, and others can be custom-written. The simplest and most ideal kind of narrow-band signal has a constant intensity and drift rate; such signals can be created straightforwardly through the wrapper function \texttt{add\_constant\_signal}.

For a pixel at $(t, f)$ in the time-frequency spectrogram, the intensity of a synthetic signal is calculated as
\begin{equation}
    I(t, f) = I_t(t) I_f(f, I_p(t)) I_{bp}(f). \label{eq:addsignal}
\end{equation}
As such, Eq. \ref{eq:addsignal} is computed for every pixel in the spectrogram, since there is no robust way to constrain arbitrary intensity profiles. For example, even an ideal Gaussian function is non-zero at all distances and defining a suitable range depends on the experiment. For large spectrograms, it can be inefficient to calculate intensities for pixels far from the main signal, so users can provide a custom frequency range to limit the signal calculation. 

The signal calculation is fully heuristic, in that the calculation is completely user-specified and does not take other effects into account, such as FFT leakage or spectral responses. Since intensity is treated as a function of time and frequency, this process can overlook how intensities are integrated in reality. As a partial solution, \texttt{add\_signal} provides the option to separately sub-integrate within each pixel in time and frequency directions. 

In a similar vein, a difficult effect to handle robustly is Doppler smearing, in which a highly drifting signal will have its power spread into multiple frequency channels within the same time channel \citep{sheikh2019choosing}. While an analytical form exists for the spectral profile of a linearly drifting cosine signal, the smearing effect will naturally apply to more complex signals. Variable spectral profiles are not yet supported in \setigen, but from a user standpoint, it would be tedious to manually construct custom smearing profiles that change at each timestep. Using a similar process to numerical integration, \texttt{add\_signal} has the option to approximate Doppler smearing by computing and averaging a given number of copies of the signal, spaced evenly between signal center frequencies in adjacent timesteps. For instance, for the $i$th time channel at $t=t_{i}$, copies of the signal centered at even spacings between $I_p(t_{i})$ and $I_p(t_{i+1})$ are averaged together to get the $i$th spectral profile. This is done for all time channels, so that channels with smaller signal drifts will be brighter than those with larger signal drifts by the correct ratio, as long as the number of copies gives enough coverage over the channel with the largest signal drift. 

Sometimes it can be difficult or unwieldy to wrap up a desired signal property into a separate function, or perhaps there is existing external code that produces such properties. In these cases, we can instead use NumPy arrays to describe these signals, rather than functions. As of now, the \texttt{path}, \texttt{t\_profile}, and \texttt{bp\_profile} arguments can be arrays.

\subsubsection{Common Frame Operations}

\begin{figure*}
\begin{center}
  \includegraphics[width=\textwidth]{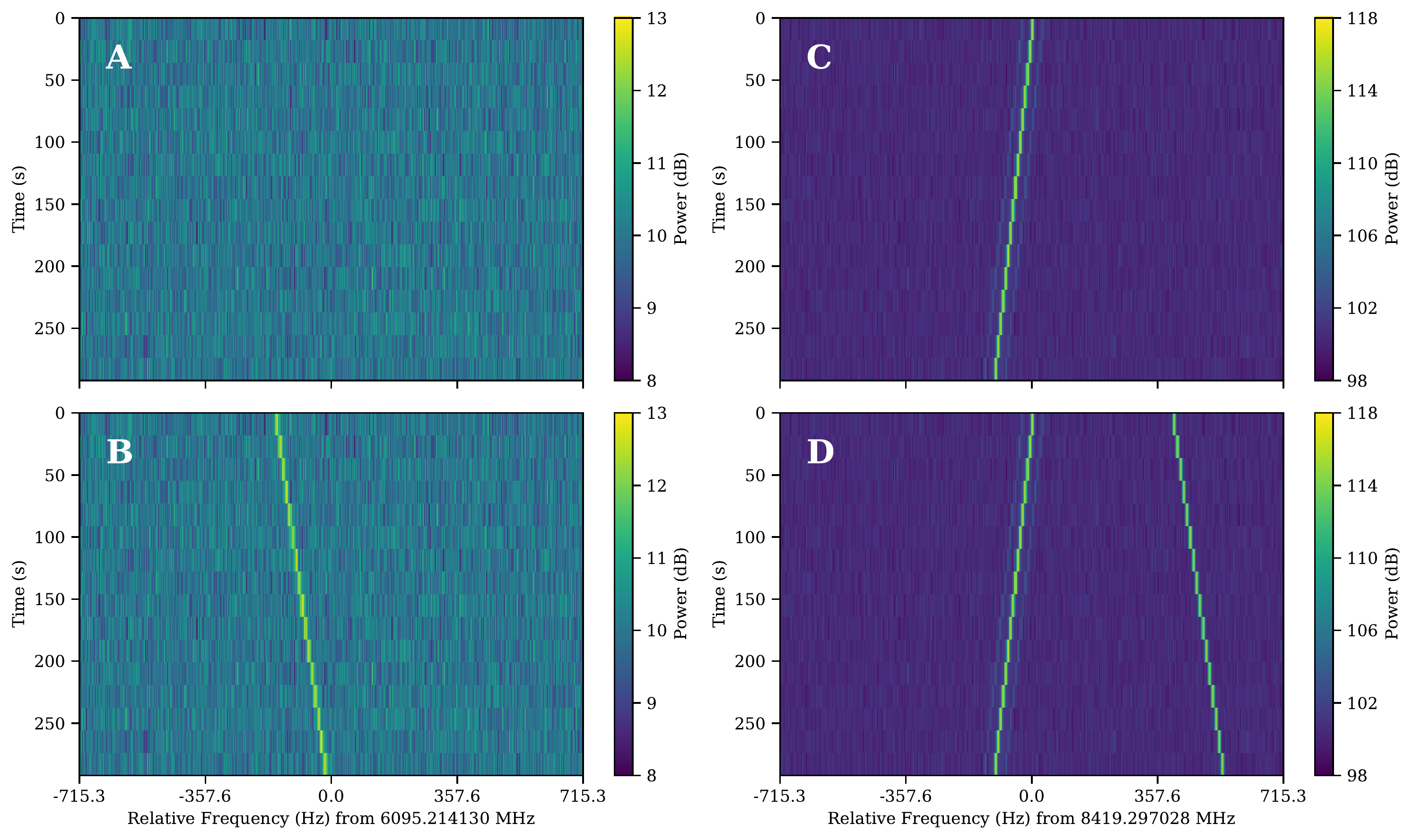}
  \caption{Radio spectrogram plots created from \setigen frames. \textbf{A}: Frame with only synthetic chi-squared noise. \textbf{B}: Frame from panel A with an injected synthetic signal at SNR=30. \textbf{C}: ``Real'' GBT observation of Voyager I carrier signal at X-band. \textbf{D}: Frame from panel C with an injected synthetic signal at SNR=1000, with the same drift rate as the injected signal in panel B.}
  \label{fig:spec-multiplot}
\end{center}
\end{figure*}

Besides supporting noise and narrow-band signal injection, \setigen comes with a set of tools for radio spectrogram analysis. These range from convenience functions for parameter calculations to frame-level data transformations. 

For instance, estimating the SNR of a signal in an integrated spectrum is a common step in radio analysis. This can be done through a frame's \texttt{integrate} function, which can also be used along the frequency axis to produce an intensity time series array.

To inject a signal at a desired SNR, the \texttt{get\_intensity} function calculates the requisite signal level as
\begin{equation}
    I_t = \text{SNR} \cdot \frac{\sigma_b}{N_t^{1/2}},
\end{equation}
assuming that the frame has background noise with standard deviation $\sigma_b$ and that the SNR is measured by dividing the integrated signal maximum by the integrated noise deviation. As discussed in Section \ref{subsubsec:noise}, each frame tracks an estimate of $\sigma_b$ calculated using iterative sigma clipping and updates it when synthetic noise is injected.

It can be convenient to define signals in terms of the pixels they traverse rather than the frequencies. To convert between these for a given frame, one can use the \texttt{get\_frequency} and \texttt{get\_index} functions. We define the \textit{unit drift rate} for a given spectrogram resolution to be the drift rate given by
\begin{equation}
    \dot{\nu}_1 = \frac{\Delta f}{\Delta t},
\end{equation}
which can be accessed with the \texttt{unit\_drift\_rate} attribute. For a linearly-drifting signal passing through the top and bottom of the frame, the corresponding drift rate can be calculated using the \texttt{get\_drift\_rate} function.

Given a frame with a linearly-drifting signal, we can ``de-drift" the frame using \texttt{setigen.dedrift}. This shifts each spectrum an appropriate amount along the frequency direction so that such a signal would, on average, appear to have zero frequency drift, making it simpler to calculate the SNR. In practice, empirical drift rates are not generally multiples of the unit drift rate, so de-drifted signals will not be perfectly aligned.

We can create a ``slice" of a frame by specifying left and right frequency indices, analogous to NumPy array slicing, by using the frame's \texttt{get\_slice} function. This results in a new frame with a truncated range, which can be helpful for isolating signals in time-frequency space for further analysis.

If one is interfacing with other BL or astronomy codebases, outputting \setigen frames to filterbank or HDF5 format can be very useful. These are done via the \texttt{save\_fil} and \texttt{save\_hdf5} functions. Frame objects can also be written and loaded with \texttt{pickle}, a convenient serialization method that can keep data and user-provided metadata together. 

\subsubsection{Demonstration: Spectrogram Module}

We present a minimal working example of creating a data frame with synthetic noise and a drifting signal. First, we construct an empty frame with the desired resolution; here, we use parameters that match those of BL's high frequency resolution data product:

\begin{minted}{python}
from astropy import units as u
import setigen as stg

frame = stg.Frame(fchans=256,
                  tchans=16,
                  df=2.7939677238464355*u.Hz,
                  dt=18.253611008*u.s,
                  fch1=6095.214842353016*u.MHz)
\end{minted}

Then, we add chi-squared noise with a desired mean, such as 10:
              
\begin{minted}{python}    
frame.add_noise(x_mean=10, noise_type='chi2')
\end{minted}

Finally, we add a simple drifting signal through our frame at SNR=30 and plot the result in decibels (dB). The inputs to \texttt{add\_signal} shown below are pre-written library functions that themselves return the functions described in Section \ref{subsubsec:signal}. Since they are indeed Python functions by type, the signal parameters allow for much more flexibility beyond this basic example.
              
\begin{minted}{python}
frame.add_signal(
    stg.constant_path(
        f_start=frame.get_frequency(index=100),
        drift_rate=2*u.Hz/u.s
    ),
    stg.constant_t_profile(
        level=frame.get_intensity(snr=30)
    ),
    stg.gaussian_f_profile(width=10*u.Hz),
    stg.constant_bp_profile(level=1)
)

frame.bl_plot()
\end{minted}

The frames after adding noise and after adding the signal are shown in Figures \subplotref{fig:spec-multiplot}{A} and \subplotref{fig:spec-multiplot}{B}. 

We also show an example with a signal detected from Voyager I in an X-band observation using the GBT, in Figure \subplotref{fig:spec-multiplot}{C}. Injecting a signal into the Voyager frame with the same drift rate as in the example (Figure \subplotref{fig:spec-multiplot}{B}), now at SNR=1000, we get Figure \subplotref{fig:spec-multiplot}{D}.

\subsection{Raw Voltage Module}
\label{subsec:voltage}

\begin{figure*}
\begin{center}
  \includegraphics[width=\textwidth]{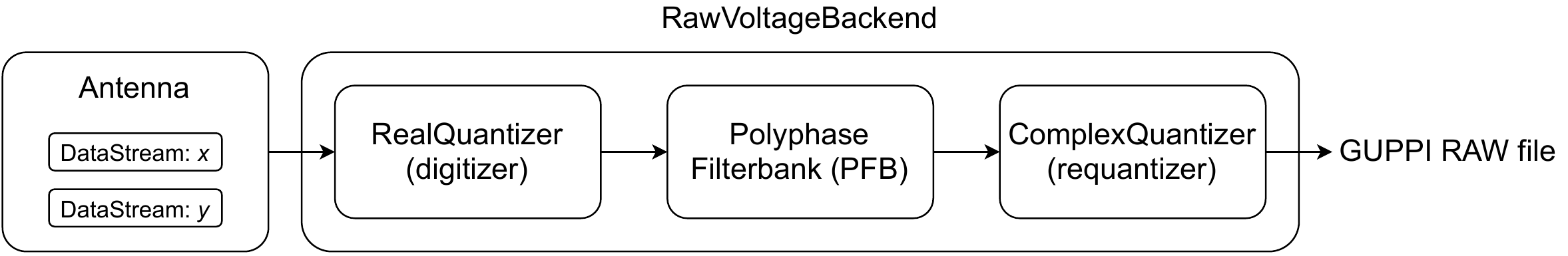}
  \caption{Basic layout of a voltage pipeline written using \texttt{setigen.voltage}.}
  \label{fig:voltage}
\end{center}
\end{figure*}

The raw voltage module is designed for synthesizing complex voltage data, providing a set of classes that models the signal processing pipeline described in Section \ref{sec:signalchain}. Instead of directly synthesizing spectrogram data, we can produce real voltages, pass them through a virtual pipeline, and record to file in GUPPI raw format. As this process models actual hardware used by BL for recording raw voltages, this enables lower level testing and experimentation.

The basic signal flow is shown in Figure \ref{fig:voltage}. At the lowest level, a \texttt{DataStream} can accept noise and signal sources (as Python functions) and generate real voltages on demand. An \texttt{Antenna} models an antenna or dish used in radio telescopes and has one or two \texttt{DataStream} objects, corresponding to linear polarizations that are unique and not necessarily correlated. As described in Section \ref{sec:signalchain}, the sampled real voltages are passed to a processing pipeline which consists, at its core, of a digitizer, a polyphase filterbank (PFB), and a requantizer. In hardware, processing is done in fixed point arithmetic on an FPGA, but for simplicity, we use floating point. The digitizer quantizes input voltages to a specified number of bits and a target full width at half maximum (FWHM) in the quantized voltage space. The filterbank implements a software PFB, coarsely channelizing input voltages. The requantizer takes the resulting complex voltages and quantizes each component to either 8 or 4 bits, suitable for saving into GUPPI raw format. 

The \texttt{RawVoltageBackend} object wraps around these elements and connects the full pipeline together. Given an observation length in seconds or a number of data recording ``blocks,'' the main function \texttt{record} retrieves real voltage samples as needed and passes them through each backend element, finally saving the quantized complex voltages out to disk.

Since voltage data is taken with very high sample rates, e.g. Gigasamples/sec (Gsps), the voltage module is much more computationally expensive than the spectrogram module. To increase efficiency, most of the data manipulations are done with matrix operations, allowing for GPU acceleration with \texttt{CuPy} \citep{cupy_learningsys2017}.

\subsubsection{Antennas and DataStreams}

The \texttt{DataStream} class represents a stream of real voltage data for a single polarization and antenna. A data stream has an associated sample rate $f_s$, such as 3\,GHz for the BL DR. As of now, the voltage module does not implement heterodyne mixing or bandpass filtering. Instead, data streams use a reference frequency \texttt{fch1} and frequency sign (ascending or descending from \texttt{fch1}) for voltage calculations. 

The \texttt{Antenna} class is similarly defined by a sample rate, reference frequency, and frequency sign. For two linear polarizations, an \texttt{Antenna}’s data streams are available via the \texttt{x} and \texttt{y} attributes. For one polarization, only the former is available. For convenience, the \texttt{streams} attribute gets the list of available data streams for an antenna. One can add noise and signal sources to these individual data streams. 

Real voltage noise is modeled as ideal Gaussian noise and added through the \texttt{add\_noise} function. Note that this actually stores a Python function to the data stream that is only evaluated when \texttt{get\_samples} is called. It also updates the data stream's \texttt{noise\_std} attribute, which keeps track of the standard deviation of the voltages in that data stream. This is useful for injecting signals at target spectrogram SNRs. 

Drifting cosine signals can be added to a data stream using \texttt{add\_constant\_signal}. For more complex signals, one can write custom voltage functions to add using \texttt{add\_signal}. Voltage signal sources are Python functions that accept an array of timestamps and output a corresponding sequence of real voltages. Here is a simple example that adds a non-drifting cosine signal with frequency \texttt{f\_start}:

\begin{minted}{python}
def cosine_signal(ts):
    delta_f = f_start - antenna.x.fch1
    return np.cos(2 * np.pi * delta_f * ts)

antenna.x.add_signal(cosine_signal)
\end{minted}

As custom signals are added, the \texttt{noise\_std} parameter may no longer accurately reflect the background noise. In these cases, one can run the data stream's \texttt{update\_noise} function to estimate noise empirically. This is not done by default to save computation, especially when there are multiple well-behaved voltage sources (e.g. Gaussian noise, cosine signals).

\subsubsection{Quantization}

The quantization process takes a continuous input voltage distribution and scales it to a target distribution that can be described by $N_{\text{bits}}$ bits. Since real voltage noise can be modeled by a Gaussian process, we can define this scaling in terms of the standard deviation or FWHM. 

For real voltages $\{v\}$, target bit size $N_{\text{bits}}$, target mean $\mu_q$ (ideally 0), and target standard deviation $\sigma_q$, the quantized voltages $v_q$ are given by:
\begin{align}
    v_s &= \left\lfloor \frac{\sigma_q}{\sigma_v} (v - \langle v \rangle) + \mu_q \right\rfloor \\
    v_q &= \min(\max(-2^{N_{\text{bits}} - 1}, v_s), 2^{N_{\text{bits}} - 1} - 1)
\end{align}
We can define quantizers in terms of a target FWHM $w_q$, in which case $\sigma_q = \frac{w_q}{2\sqrt{2\ln 2}}$.

The digitizer quantizes real voltages, while the requantizer receives complex voltages and quantizes per complex component. Quantization is run per polarization and antenna, and background statistics can be cached to save computation in subsequent calls. This is facilitated by the \texttt{RealQuantizer} and \texttt{ComplexQuantizer} classes.

\subsubsection{Polyphase Filterbank}

The \texttt{PolyphaseFilterbank} class implements and applies a PFB to quantized input voltages. Instead of directly applying a $P$-point FFT, a PFB first splits incoming voltages between $P$ branches and lets $M$ samples accumulate in each branch \citep{price2021spectrometers}. A windowing function is applied over the $M\times P$ samples, the samples are summed over the $M$ so-called polyphase taps, and finally a $P$-point FFT is taken of the result to get complex raw voltages in $N_{\text{coarse}}=P/2$ coarse channels. Further samples are read in groups of $P$ and split between the PFB branches; accumulated samples step forward to the next tap to make room. PFBs have a better channel response than standard FFTs, especially as $M$ increases, and are common in high spectral resolution radio backends \citep{price2021spectrometers}. 

The two main parameters for a \texttt{PolyphaseFilterbank} are the number of taps $M$ and the number of branches $P$. Since the PFB works on $MP$ samples at once, the object continuously caches samples for on-demand computation. The PFB also accepts a symmetric windowing function as an argument (Hamming, by default) and generates $MP$ coefficients up front \citep{blackman1958measurement}.

\subsubsection{Combining Components and Recording Data}

The \texttt{RawVoltageBackend} class contains the full machinery to collect, process, and write complex voltage data to GUPPI raw files, as in the standard pipeline shown in Figure \ref{fig:voltage}. Nevertheless, since the individual signal processing components are all exposed as part of the voltage module, custom pipelines can be written by chaining them in different ways. 

A \texttt{RawVoltageBackend} takes in components external to the data recording process as parameters, such as the antenna, digitizer, PFB, and requantizer. All other parameters and functions are specific to data recording and actually obtaining data from the external components. 

As described by \cite{Lebofsky:2019}, the block size $N_{\text{blocksize}}$ refers to the number of bytes in a single block of data in GUPPI format. Each data block has an associated header with observing metadata, such as target and frequency information. The number of blocks per file also must be specified to size individual raw files; multiple raw files may be associated with a single pointing. For standard 5 minute GBT observations, BL DR uses $N_{\text{blocksize}}=134217728$ with 128 blocks per file.

To specify the coarse channels that are actually recorded to disk, we can set the starting index and the number of consecutive channels $N_{\text{chan}}$ to ultimately save. Purely for computational efficiency, we always perform a full FFT and truncate to obtain the desired coarse channels, instead of directly doing the transform operation on the subset of coarse channels. Especially when using a GPU to accelerate synthesis, this can fill up memory rather quickly, potentially to the point of overflow. Therefore, the \texttt{RawVoltageBackend} has an additional option to divide individual data blocks into a given number of sub-blocks, such that each sub-block will fully fit in memory.

For a single antenna, the number of bytes $N_{\text{blocksize}}$ in a block can be related to the number of time channels $N_{t,\text{block}}$ corresponding to a single block in (non-integrated) spectrogram format as

\begin{align}
    N_{\text{blocksize}} &= 2 N_{\text{pol}} \left( \frac{N_{\text{bits,r}}}{8} \right) N_{\text{chan}} N_{t,\text{block}} \\
    &= \frac{1}{4} N_{\text{pol}} N_{\text{bits,r}} N_{\text{chan}} N_{t,\text{block}},
\end{align}
based on the structure of raw files as described by \cite{Lebofsky:2019}.

\subsubsection{Multi-Antenna Support}

To simulate voltage data for interferometric pipelines, it can be useful to synthesize raw voltage data from multiple antennas. \setigen supports synthesizing multi-antenna output through the \texttt{MultiAntennaArray} class, which creates a list of $N_{\texttt{ant}}$ antennas each with an associated integer delay (in time samples). In addition to the individual data streams that allow the user to add noise and signals to each antenna, there are ``background" data streams \texttt{bg\_x} and \texttt{bg\_y} in \texttt{MultiAntennaArray}, representing correlated noise or RFI that is detected at each antenna, subject to the (relative) delays. Signals and noise can therefore be added to the background across all array elements as well as to individual antennas.

The only difference in the pipeline is instead of supplying a \texttt{Antenna} as input to a \texttt{RawVoltageBackend}, one would supply a \texttt{MultiAntennaArray}. Then, the output is saved as a multi-antenna extension of the GUPPI raw format. 

\subsubsection{Creating Signals at a Target Spectrogram SNR}

During the course of the full signal processing pipeline, an injected cosine signal passes through multiple quantization and FFT steps. In many SETI experiments, a signal's SNR in spectrogram data is used for thresholding and analysis, so it is important to be able to estimate this SNR given pipeline parameters. 

Suppose that we have a cosine signal with amplitude $A$ at a frequency corresponding to the center of a fine spectral channel, and that this signal is injected onto a background of Gaussian noise $\mathcal{N}(0, \sigma_v^2)$. Since the voltage data is real-valued, the signal magnitude becomes $A/2$ in frequency space. As the voltages pass through the coarse and fine channelization steps, the signal magnitude picks up factors of $P$ and $N_{\text{fine}}$, respectively, compared to the background noise. 

The background noise will follow a chi-squared distribution with $\text{DOF} = 2 N_{\text{pol}} N_{\text{int}}$ (Section \ref{subsubsec:noise}), scaled by multiplicative factors arising from quantization and FFT calculations. Since the input voltage noise has variance $\sigma_v^2$, the standard deviation of the noise power $\sigma_b$ will be proportional to the standard deviation $\sigma_{b,0}$ of a chi-squared distribution with mean $\sigma_v^2$. The time integration step to get the SNR will reduce this noise by a factor of $N_t^{1/2}$.

To get an expression for $N_t$ given observation parameters, suppose our synthetic observation has $N_{\text{block}}$ total blocks and that the time covered by a single block is $\tau_{\text{block}}$. Then, we have the following equations:
\begin{align}
    \Delta t &= \frac{N_{\text{int}}}{\Delta f} = \frac{P}{f_s} N_{\text{fine}} N_{\text{int}} \\
    \tau_{\text{block}} &= N_{t,\text{block}} \Delta t \\
    N_t &= \frac{N_{\text{block}} \tau_{\text{block}} }{N_{\text{int}} \Delta t} = \frac{N_{\text{block}} N_{t,\text{block}}}{N_{\text{int}}}.
\end{align}

Combining all of these factors, we can express the final SNR of the signal as the ratio between the integrated (mean) signal power and the integrated background noise standard deviation as
\begin{align}
    \sigma_{b,0} &= \sigma_v^2 \left(\frac{2}{\text{DOF}}\right)^{1/2} \\
    \text{SNR} &= \frac{I}{\sigma_b} = \frac{(A/2)^2 P N_{\text{fine}}}{\sigma_{b,0}/N_t^{1/2}}.
\end{align}
This yields the amplitude or signal level in terms of the target SNR:
\begin{equation}
    A = \left( \text{SNR} \cdot \frac{ 4\sigma_{b,0} }{ P N_{\text{fine}} N_t^{1/2}}   \right)^{1/2}
\end{equation}
Notice that $A$ has a linear dependence on the standard deviation $\sigma_v$ of the real voltage noise in a data stream, which can arise from multiple sources, especially in a multi-antenna array. Given pipeline parameters, the \texttt{get\_level} function can be used to calculate $A/\sigma_v$.

For a non-drifting cosine signal, we can also approximate the effect of spectral leakage between fine channels by comparing the signal frequency to the nearest channel central frequency. A signal with amplitude $A$ centered at a frequency $\delta f$ away from the center of the closest fine spectral channel will have its power $I$ attenuated by\footnote{$\text{sinc } x = \sin{x}/x$}
\begin{equation} \label{eq:leakage}
    \frac{I'}{I} = \text{sinc}^2 \left( \frac{|\delta f|}{\Delta f} \right).
\end{equation}
Since intensity goes as voltage squared, we provide a function \texttt{get\_leakage\_factor} to calculate an amplitude adjustment factor $f_l$ to easily scale from $A$ to a new amplitude $A'$ that corresponds to the non-attenuated intensity:
\begin{align}
    f_l &= \frac{1}{\text{sinc} \left( \frac{|\delta f|}{\Delta f} \right)} \\
    A' &= f_l A.
\end{align}

Finally, for a linearly-drifting cosine signal, if the drift rate $\dot{\nu}$ exceeds the unit drift rate $\dot{\nu}_1$, signal power will be smeared across multiple frequency bins in spectrogram data. This is a linear effect in spectrogram data, so cosine amplitudes should be increased by a factor of $(\dot{\nu}/\dot{\nu}_1)^{1/2}$ to counter-act the apparent loss in power.

\subsubsection{Injecting Synthetic Signals into Raw Voltage Data}

In addition to creating fully synthetic complex voltage data from scratch, the \texttt{RawVoltageBackend} supports injecting or adding synthetic data to existing observational GUPPI raw data. The pipeline remains mostly the same, except for a few important differences that we detail below.

In order to get meaningful results, we must know and match details about the specific signal processing pipeline that produced the existing raw data. \setigen provides a helper function called \texttt{get\_raw\_params} to extract header information from the raw data file, but other information must be provided separately by the user, such as the sampling rate and PFB parameters.

Since recorded voltage data has already gone through multiple quantization steps, we cannot directly add time series voltages together (i.e. at the original ADC sampling rate). Instead, we choose to synthesize complex voltage data separately, add it to the recorded voltage data, and apply a final quantization step to match the initial distribution as best as possible. 

However, this process requires that we create and process signals that are not necessarily embedded in noise. In typical narrow-band signal injection scenarios, we wish to synthesize and inject signals whose distributions are non-Gaussian (e.g. a cosine signal). Since the quantization steps assume that the input and output voltage distributions are both Gaussian, attempting to quantize bare narrow-band signals will cause distortion and introduce clipping artifacts. Furthermore, without a reference noise distribution, quantization can scale the magnitude of processed signals in undesired ways, making SNR estimation difficult. 

To address these issues, we approach the quantization steps differently. If there is already a synthetic noise source, we proceed normally through all steps in the pipeline. Otherwise, we skip the initial digitization step before the PFB, and instead treat the input voltages as if they followed a zero-mean Gaussian distribution with variance 1. Using a reference distribution allows us to set signal magnitudes with the \texttt{get\_level} function to achieve target SNR levels. We then estimate the post-PFB mean and standard deviation of the reference Gaussian voltages and quantize the synthetic voltages based on these values instead of those from the ``real’’ synthetic distribution. This way, if the synthesized voltages were actually embedded in $\mathcal{N}(0, 1)$ noise, the resulting signal quantization would be very similar.

For each data block in the recorded raw file, the \texttt{RawVoltageBackend} will set requantizer statistics (target mean $\mu_q$ and target standard deviation $\sigma_q$) calculated from the existing data for each combination of antenna, polarization, and complex component. The synthetic voltages are requantized to the corresponding standard deviations in each complex component, but instead of centering to the target mean, they are centered to zero mean. This is so that when we add the quantized synthetic data to the existing data, we do not change the overage voltage mean. After these are added together, we finally requantize once more to the target mean and target standard deviation to match the existing data statistics and magnitudes as best as possible.

\subsubsection{Demonstration: Voltage Module}

\begin{figure}
\begin{center}
  \includegraphics[width=.45\textwidth]{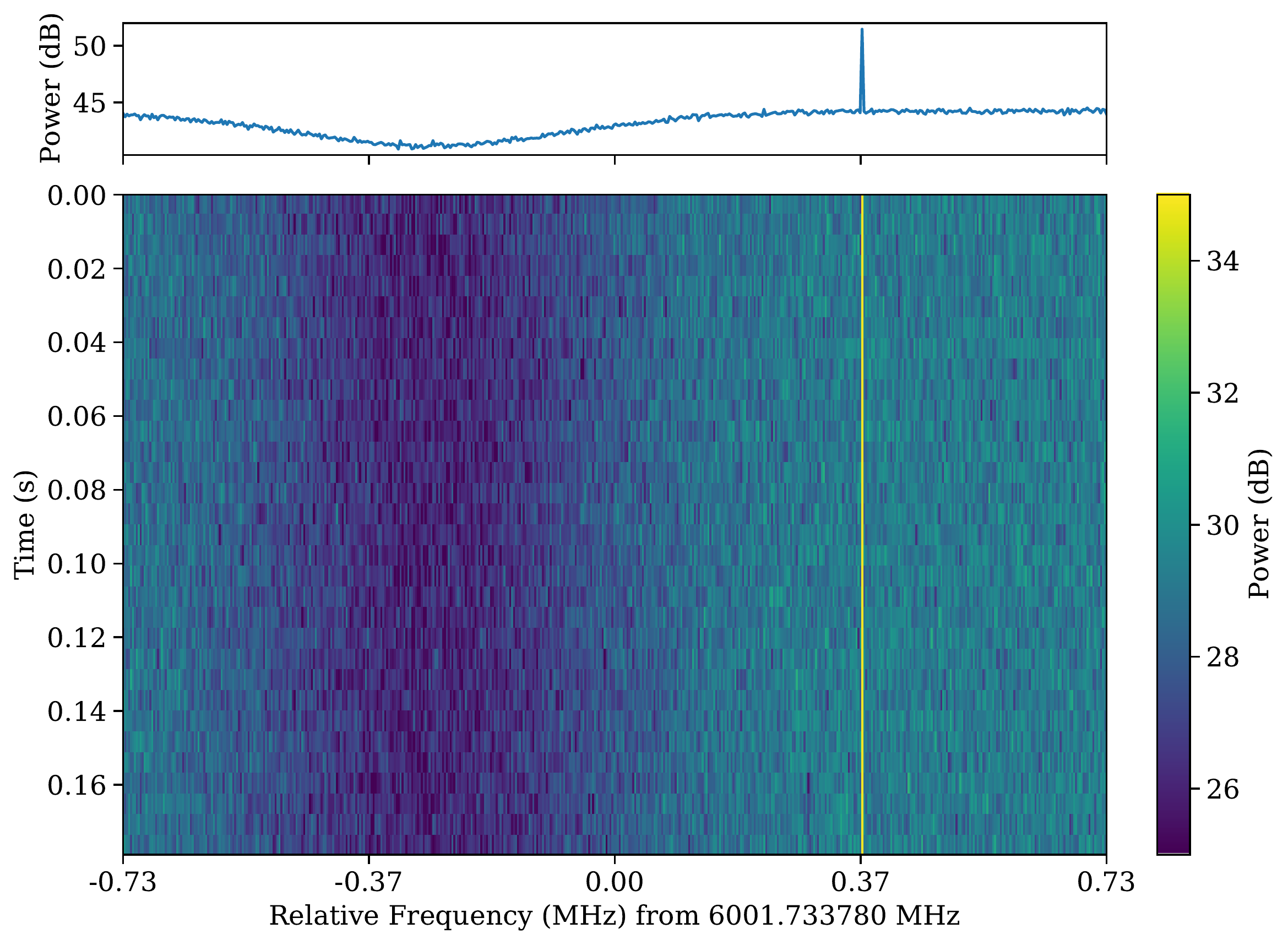}
  \caption{Spectrogram derived from synthetic raw voltages, showing the edge of the coarse channel bandpass shape and a bright, slightly drifting cosine signal. The top panel shows an integrated profile, showing PFB scalloping loss towards the left and the synthetic signal towards the right.}
  \label{fig:voltagespectrogram}
\end{center}
\end{figure}

Here, we present a simple pipeline created with the raw voltage module to inject a drifting cosine signal in Gaussian noise. First, we create the signal processing elements:

\begin{minted}{python}
from astropy import units as u
from setigen.voltage import *

d = RealQuantizer(target_fwhm=32,
                  num_bits=8)

f = PolyphaseFilterbank(num_taps=8,
                        num_branches=1024)

r = ComplexQuantizer(target_fwhm=32,
                     num_bits=8)
\end{minted}

Then, we create the antenna, setting the sampling rate and reference frequency. With two polarizations, we can add Gaussian noise and a constant amplitude, Doppler drifting cosine signal to both data streams:

\begin{minted}{python}
a = Antenna(sample_rate=3*u.GHz,
            fch1=6000*u.MHz,
            ascending=True,
            num_pols=2)

for s in a.streams:
    s.add_noise(v_mean=0,
                v_std=1)

    s.add_constant_signal(f_start=6002.1*u.MHz,
                          drift_rate=-2*u.Hz/u.s,
                          level=0.004)
\end{minted}

We connect these components through the recording backend, defining the dimensions and size of the final raw voltage data product, and record a block of data to file.

\begin{minted}{python}
rvb = RawVoltageBackend(a,
                        digitizer=d,
                        filterbank=f,
                        requantizer=r,
                        start_chan=0,
                        num_chans=64,
                        block_size=134217728,
                        blocks_per_file=128,
                        num_subblocks=32)

rvb.record(output_file_stem='example_1block',
           num_blocks=1,
           length_mode='num_blocks',
           header_dict={'TELESCOP': 'GBT'},
           verbose=True)
\end{minted}

After saving the raw voltages to disk, we reduce using \texttt{rawspec} with $N_{\text{fine}}=1024$ and $N_{\text{int}}=4$. A snippet of the resulting spectrogram output is shown in Figure \ref{fig:voltagespectrogram}, where intensities are plotted on a decibel scale. The signal is readily apparent, as is the frequency bandpass shape arising from the PFB.

\section{Discussion}
\label{sec:discussion}

\subsection{Limitations}

While \setigen is a flexible library that enables quick narrow-band dataset generation, it is important to discuss the limitations when using it for science. 

First and foremost, \setigen relies on heuristic, user-defined signals, rather than simulations from first principles. The search for technosignatures is necessarily informed by human bias, specifically applied via our assumptions about a technosignature's potential characteristics and morphology. It is possible that radiation from an extraterrestrial intelligence will exist in a form that we have not considered or designed searches for. Even when we consider only the problem of excision of anthropogenic RFI, we have to be careful when applying algorithms developed using the simplest of narrow-band signals. Although there might never be a way of fully emulating the breadth and variety of the RFI environment, \setigen can still be used to generate labeled, complex signals to test the efficacy of new and existing algorithms.

In a similar vein, the spectrogram module enables users to quickly generate signals that ``look'' like the narrow-band signals we see in observations. However, since spectrogram signal injection does not have access to phase information, it is impossible to replicate the ``correct'' intensity statistics when adding a signal to integrated Stokes I noise. For example, adding a perfect cosine signal to zero-mean Gaussian noise in the voltage domain results in a non-central chi-squared intensity distribution in Stokes I data, but adding a signal with constant intensity directly to chi-squared noise in a spectrogram does not result in the same distribution \citep[over the pixels occupied by that signal;][]{mcdonough1995detection}. While this effect is negligible for high SNR signals, algorithms developed to target low SNR signals may suffer from intrinsic inaccuracies in the intensity statistics.

Signal injection in the complex voltage domain also has limitations since we are not able, in software, to directly add signals in the real (analog) voltage stage. Raw data is quantized multiple times in hardware, so the injection step has to take place using complex voltages that are quantized in a similar way. While fundamental steps in the pipeline are linear, such as PFB operations (Eq. \ref{eq:pfb}), quantization inherently breaks this linearity. Because of this, summing real and synthetic voltages that are independently processed can lead to artifacts and intensity discrepancies that would not arise if we could inject at the start of the signal processing pipeline. 

\subsection{Future Directions}

\setigen is written and developed with the needs of SETI researchers in mind, so new functionality and improvements are constantly being added. Here, we describe some potential enhancements that may be added in the near future.

As it stands, the spectrogram module is especially targeted at producing small frames with synthetic signals rather than injecting into large, broadband observations. While this suffices in many cases, it may be useful to inject within large data files in which frequency bandpass shapes significantly change the background intensities. For instance, for use in SNR estimation, \setigen calculates background noise statistics over an entire frame rather than localized around the target signal injection frequency. For a large enough frame, this is both an inefficient and inaccurate calculation due to variable bandpass shapes. An improvement would be to localize the noise calculation to a window around the target injection site, as well as to similarly localize the signal injection calculation to prevent unnecessary computation.

The spectrogram module is also currently designed expressly to synthesize narrow-band signals. There are many similarities in both signal processing and experimental design between technosignature searches and searches for time-varying phenomena such as pulsars and fast radio bursts (FRBs); \setigen could thus be expanded to include broadband signal injection \citep{2018ApJ...866..149Z, gajjar2021breakthrough}. 

An exciting potential addition is to use parameterized ML methods to create labeled, realistic signals. By taking ideas from style transfer, a synthetic RFI signal could be created by specifying heuristic parameters and having an ML model generate such a signal with RFI-like properties \citep{gatys2016image, dai2017towards}. While generative adversarial networks (GANs) have been used before to create radio spectrograms \citep{zhang2018self}, conditional GANs that accept input parameters might help produce more specific, labeled signals, which can be better for certain SETI experiments. Furthermore, better RFI modeling could help improve ML-based searches for astrophysical phenomena like FRBs in the presence of different classes of RFI.

Some of these enhancements may use a lot more computational power than the current synthesis process, so the option to GPU-accelerate the standard spectrogram module would be critical. Some of these enhancements may require a more careful look at file input/output methods when reading and writing large observational data files to avoid unnecessary or slow operations. 

The raw voltage module can also be expanded to support alternate radio telescope configurations and backends, such as those behind interferometers like MeerKAT \citep{jonas2009meerkat}. While \setigen already has basic multi-antenna functionality, it could be helpful to build on this with general-use utilities, such as routines that predict how a given injected signal would appear across multiple antennas or beams. The voltage module could also support additional requantization and recording modes, such as 2 and 16-bit. As interferometer usage in modern radio SETI continues to grow, \setigen capabilities can be extended to help test signal detection in commensal and beam-formed observations \citep{czech2021meerkat}.

\section{Summary}

In this paper, we presented \setigen, an open-source Python library for the creation and injection of synthetic narrow-band radio signals. \setigen can produce both finely channelized spectrogram data and coarsely channelized complex voltage data. The spectrogram module is designed to be intuitive and quick to use to facilitate the construction of synthetic datasets for SETI experiments and testing. While the voltage module is more complex and computationally intensive, it enables analysis of signals that pass through a software-defined pipeline, which can be helpful in understanding the implications of the instrumentation pipeline itself in SETI searches. 

\setigen is constantly being improved with the needs of SETI research in mind. As open-source software, the library is freely available, and we encourage the SETI community to use and contribute to it.

\section{Acknowledgements}

Breakthrough Listen is managed by the Breakthrough Initiatives, sponsored by the Breakthrough Prize Foundation. The Green Bank Observatory is a facility of the National Science Foundation, operated under cooperative agreement by Associated Universities, Inc. We thank the staff at the Green Bank Observatory for their operational support.

\software{\texttt{NumPy} \citep{oliphant2006guide},
\texttt{CuPy} \citep{cupy_learningsys2017},
\texttt{SciPy} \citep{virtanen2020scipy},
\texttt{Astropy} \citep{robitaille2013astropy},
\texttt{Blimpy} \citep{2019JOSS....4.1554P},
\texttt{H5py} \citep{collette2017h5py},
\texttt{Matplotlib} \citep{hunter2007matplotlib}}

\bibliographystyle{aasjournal}
\bibliography{references}

\end{document}